\documentclass{iaus}
\usepackage{graphicx}

\title[Evolution of high-mass stars] 
{Evolution and chemical and dynamical effects of high-mass stars}

\author[Georges Meynet et al.]   
{Georges Meynet$^1$, Cristina Chiappini$^{1,2}$,Cyril Georgy$^1$, Marco Pignatari$^{3,4}$, 
Raphael Hirschi$^3$, Sylvia Ekstr\"om$^1$ and Andr\'e Maeder$^1$}

\affiliation{$^1$Geneva University, Geneva Observatory \\
CH-1290 Versoix, Switzerland \\ email: {\tt georges.meynet@obs.unige.ch} \\[\affilskip]
$^2$Osservatorio Astronomico di Trieste \\
Via G. B. Tiepolo 11, I - 34131 Trieste, Italia \\ email: {\tt Cristina.Chiappini@obs.unige.ch} \\[\affilskip]
$^3$Keele University, Keele \\ Staffordshire ST5 5BG, United Kingdom\\email: {\tt mpignatari@gmail.com}
\\[\affilskip]
$^4$Joint Institute for Nuclear Astrophysics \\ University of Notre Dame, Notre Dame, IN 46556, United States}

\pubyear{2008}
\volume{254}  
\pagerange{119--126}
\jname{The Galaxy Disk in Cosmological Context}
\editors{J. Andersen, J. Bland-Hawthorn \& B. Nordström eds.}
\begin{document}

\maketitle

\begin{abstract}
We review general characteristics of massive stars, present the main observable constraints that
stellar models should reproduce. We discuss the impact of massive star nucleosynthesis on the early phases of the chemical evolution of the Milky Way (MW). We show that rotating models can account for the important primary nitrogen production needed at low metallicity. Interestingly such rotating models can also better account for other features as
the variation with the metallicity of the C/O ratio. Damped Lyman Alpha (DLA) systems present similar characteristics as the halo of the MW for what concern the N/O and C/O ratios. Although in DLAs, the star formation history might be quite different from that of the halo, in these systems also, rotating stars (both massive and intermediate) probably play an important role for explaining these features. The production of primary nitrogen is accompanied by an overproduction of other elements as $^{13}$C, $^{22}$Ne and s-process elements. We show also how the observed variation with the metallicity of the number ratio of type Ibc to type II supernovae may be a consequence of the metallicity dependence of the line-driven stellar winds.
\end{abstract}

\firstsection 
              
\section{Massive stars as ``cosmic engines''}             
              
Massive stars are cosmic engines (see the recent proceedings entitled
``Massive stars as Cosmic Engines'' of the IAU Symp. 250) and thus represent a key
ingredient in the evolution of the galaxies. At first sight this might be very surprising since these objects are very rare. Indeed one counts 3 stars with initial masses above 8 M$_\odot$
for 1000 stars with masses between 0.1 and 120 M$_\odot$ (estimate based on Salpeter's IMF). They contain about 14\% of the stellar mass. 

Thus both in number and in mass, massive stars represent
small fractions.
What makes them nevertheless very important objects is that massive stars are very ``generous'' objects, injecting in the interstellar medium, in {\it short timescales} (between ~3 and 30 million years) great
amounts of radiation, mass and mechanical energy:
\begin{itemize}
\item {\bf Radiation:} using the mass-luminosity relation, a 100 M$_\odot$ has a luminosity about 1 million times higher than the Sun. The high luminosity of massive stars allows them to be observed as individual objects well beyond the Local Group. 
Of course when the star explodes as a core collapse supernova, individual events can be seen at still much greater distance. Long soft GRB are believed to be associated to core collapse supernova events. One of the farthest event of that kind presently known is at a redshift of 6.26 (Cusumano et al.~2006)! 

Even in galaxies so far away that individual stars cannot be seen in them, it is still possible
to detect in the spectrum of their integrated light signature of the presence of very massive stars undergoing strong stellar winds (Wolf-Rayet stars). The broad emission lines, which are formed in the fast expanding envelope of these stars, are sufficiently important to emerge above
the continuum produced by the background stellar populations (see Brinchmann et al. 2008 for a recent survey of galaxies with Wolf-Rayet signatures in the low-redshift Universe). 
The analysis of such features allow to infer information about the number of such stars. From narrow emission lines as H$\beta$, it is possible to infer the number of O-type stars and thus to obtain a number fraction of WR to O-type stars in the starburst regions of distant galaxies.
The WR/O ratio contains information on the star formation history, the strength of the starburst and its age. Of course the quality of the information deduced in that way depends on
the quality of the stellar models used. 

The collective effect of massive stars is of first importance to understand the photometric evolution of galaxies. Typically about 2/3 of the visible light of galaxies arises from the massive star populations. 
Their high ionizing power gives birth to HII regions which trace the regions of recent star formation. The strong UV luminosity of massive stars has also been (and is still) used to deduce the history of star formation in the Universe (see e.g. Hopkins \& Beacom 2006 for a discussion
of the cosmic star formation history).
In dusty galaxies, part of the UV light heats the dust and makes the galaxies to glow in IR (see e.g. the discussion of P\'erez-Gonz\'alez et al. 2006 on M81). Interestingly the ionizing front which expands with time around massive stars can trigger star formation in their vicinity. Examples of such behavior is seen in the Triffid nebula (see Hester et al. 2005).
The ionising flux of the Pop III stars played a key role in reionizing the early Universe
(Barkana 2006).
	
\item {\bf Mass:} either through stellar winds or at the time of the supernova explosion great amounts of mass are injected back into the interstellar medium. As seen above, about 14\% of the mass of stars formed in a stellar generation,
called M$_*$ in the following, is in the form of massive stars. Nearly all this mass is ejected back into the interstellar medium by massive stars (~13\% of M$_*$), only a very small amount (about 1\% of M$_*$) remains locked into compact remnants (neutron stars or black holes). If during the formation of a black hole, all the mass is swallowed by the black hole, smaller amounts of mass are returned. Typically, if all stars more massive than 30 M$_\odot$ follow such a scenario, then only a little more than 7\% of M$_*$ are returned. 
Part of the material returned, between 3.5 and 4.5\% of M$_*$, is under the form of new synthesized elements
and thus contributes to the chemical evolution of the galaxies and of the Universe as a whole.

\item {\bf Mechanical energy:} during its lifetime a star with an initial mass of 60 M$_\odot$ will lose through stellar winds more than 3 fourths of its initial mass through stellar winds (at solar metallicity!). Assuming a mass loss rate of
4$\times$ 10$^{-5}$ M$_\odot$ per year, a velocity of the wind of 3000 km s$^{-1}$, one obtains a mechanical luminosity equal to 30000 solar luminosities or to about 10\% the radiation luminosity of the massive star. Integrating over the WR lifetime (about half a million years), one obtains that the mechanical energy injected into the surrounding amounts to 2 $\times$ 10$^{51}$ erg, {\it i.e.} a quantity of the same order of magnitude as the energy injected by a supernova explosion.
Note that only a small fraction of this energy (and of the ionising radiation) is still in the circumstellar gas at the end of the stellar lifetime. For instance, Freyer et al. (2003), using 
a two-dimensional radiation hydrodynamics code, find that 0.4\% percent of the energy emitted by a 60 M$_\odot$
under the form of ionising radiation and of wind kinetic energy is in the circumstellar gas at the end of the stellar lifetime (the supernova injection energy is not accounted for here)
\footnote{The circumstellar gas encompasses the shocked wind and photoionized HII regions, i.e. the mass inside a sphere with a radius of about 50 pc}. From this fraction 32.5\% is kinetic energy of bulk motion, 45\% is thermal energy and the remaining 22.5\% is ionization energy of hydrogen. Freyer et al. (2006) performed a similar computation for a 35 M$_\odot$. They obtain that, at the end of the stellar lifetime, 1\% of the energy released as Lyman continuum radiation and stellar wind has been transferred to the circumstellar gas. From that fraction 10\% is kinetic energy of bulk motion, 36\% is thermal energy and the remaining 54\% is ionization energy of hydrogen. 
Freyer et al. (2006) conclude that it is necessary to
consider both ionizing radiation and stellar winds for describing the interaction of OB stars with their circumstellar environment.

Collective effects of stellar winds and supernova explosion may trigger in certain circumstances galactic superwinds (see the review by Veilleux et al. 2005). These galactic winds are also loaded in new chemical species and participate to the enrichment of the intergalactic medium or intra galactic medium of clusters of galaxies.
\end{itemize}        

\begin{table}
  \begin{center}
  \caption{Overview of current grids of stellar models with rotation and mass loss.}
  \label{tab1}
 {\scriptsize
  \begin{tabular}{|l|c|c|c|c|}\hline 
{\bf Z} & {\bf Mass} & {\bf $\upsilon_{\rm ini}$} & {\bf Magn. Field} & {\bf Reference} \\ 
        &  M$_\odot$ & km s$^{-1}$                &                   &                 \\ 
 \hline
0.0 & 9,15,25,40,60,85,200 & 0 \& 800 & No & Ekstr\"om et al. (2008b) \\
    & 3,9,20,60            & 40 - 1400& No & Ekstr\"om et al. (2008a) \\
 \hline
10$^{-8}$ & 9,20,40,60,85 & 0 \& 500 - 800 & No & Hirschi (2007) \\
 \hline
10$^{-5}$ & 2,3,5,7,9,15,20,40,60 & 0 \& 300   & No  & Meynet \& Maeder (2002) \\
          & 20,30,40,50,60        & 230 - 600  & Yes & Yoon \& Langer (2005)   \\
          & 12,16,20,25,40,60     & 0 - 940    & Yes & Yoon et al. (2006)      \\
          & 3,9,20,60             & 40 - 1000  & No  & Ekstr\"om et al. (2008a) \\
 \hline
0.0005    & 20,40,60,120,200      & 0, 600, 800& No  & Decressin et al. (2007) \\
 \hline
0.001     & 20,40,60              & 230 -600   & Yes  & Yoon \& Langer (2005) \\
          & 12,16,20,25,30,40,60  & 0 - 750    & Yes  & Yoon et al. (2006)      \\ 
 \hline
0.002     & 3,9,20,60             &30 - 880   & No  & Ekstr\"om et al. (2008a) \\
          & 12,16,20,25,30,40,60  & 0 - 650    & Yes  & Yoon et al. (2006)      \\
 \hline
0.004     & 9,12,15,20,25,40,60   & 0 \& 300   & No  & Maeder \& Meynet (2001) \\
          & 30,40,60,120          & 300        & No  & Meynet \& Maeder (2005)  \\ 
          & 12,16,20,25,30,40,60  & 0 - 500    & Yes  & Yoon et al. (2006)      \\
 \hline
0.008     & 30,40,60,120          & 300        & No  & Meynet \& Maeder (2005)  \\
 \hline
0.020     & 8,10,12,15,20,25      & 0 - 470    & No  & Heger \& Langer (2000)  \\
          & 8,10,12,15,20,25      &  200       & No  & Heger et al. (2000)  \\ 
          & 9,12,15,20,25,40,60,120& 0 \& 300   & No & Meynet \& Maeder (2000) \\  
          & 9,12,15,20,25,40,60,85,120& 0 \& 300   & No & Meynet \& Maeder (2003) \\ 
          & 12,15,20,25,40,60     &0 \& 300   & No & Hirschi et al. (2004) \\
          & 12,15,20,25,35        & 200       & Yes \& No & Heger et al. (2005) \\
          & 16, 30, 40            & 210 - 560 & Yes  & Yoon et al. (2006)      \\
          & 3,9,20,60             & 30 -730   & No   & Ekstr\"om et al. (2008a) \\
\hline
0.040     & 20,25,40,60,85,120    & 0 \& 300   & No  & Meynet \& Maeder (2005)   \\          
  \hline
  \end{tabular}
  }
 \end{center}
\end{table}     
              
\section{Observed constraints for massive star models at various metallicities}  

To have a complete view of the impact of the evolution of stars on the evolution of a galaxy, models spanning the whole range of metallicities from Z=0.0 (Pop III stellar models) up to the highest metallicities observed in the considered galaxy have to be used. Different
grids of stellar models are available in the literature accounting for various effects as overshooting, mass loss,
rotation, magnetic fields, binary interactions... A subsample of recent grids of models for single stars accounting for the effects of mass loss, rotation and overshooting are presented in Table 1\footnote{Some of them include magnetic field effects according to the dynamo theory proposed by Spruit (2002).}. To use results of stellar models with some confidence
a necessary prerequisite is to check that they are able to
reproduce well observable constraints. Among the most important constraints, let us cite the following ones:
\begin{itemize}
\item {\bf Surface chemical composition of massive stars:} Many observations  
show that the surface chemical abundances of massive stars present variations during evolutionary stages and in mass domain where such changes are
not predicted by standard models\footnote{Standard models are considered here to be models in which chemical mixing occur only in convective regions.} (see for instance the recent review by Meynet et al. 2008, which summarizes the results of recent large massive star surveys in the Milky Way and the Magellanic Clouds). Understanding such changes are not only interesting for improving stellar physics, they are a key element to obtain reliable predictions for the evolutionary tracks and the chemical yields. These surface abundances are the sign that  
the chemical abundances in the stellar interiors are
different from those expected from standard models. Since what governs the evolution of a star is its internal changes of chemical composition, one can easily understand that the process responsible for these changes of the surface abundances has strong impact also on 
all the outputs of stellar models.
Explanations for these changes of the surface abundances have been proposed in the literature: rotational mixing (Heger \& Langer 2000; Meynet \& Maeder 2000) and/or mass transfer in close binary systems (Langer et al. 2008) are the two most important  invoked processes.
According to Hunter et al (2008) about 60\% of the observed B-type stars present observed characteristics compatible with the rotational mixing theory.   
\item {\bf Populations of Wolf-Rayet stars:} as is well known Wolf-Rayet (WR) stars are evaporating stars losing during their lifetime a large fraction of their initial stellar mass under the form of strong stellar winds (see the recent review by Crowther 2007). Observation shows that the number fraction of WR to O-type stars increases with the metallicity. The number ratio of WN (WR with He and N emission lines) to WC (WR with C emission lines) star also increases with the metallicity.
Such features have to be explained by stellar models. They depend on both the physics of mixing in stellar interiors and the effects of mass loss. 
Reasonable agreement between models and observations can be obtained when the effects of rotational mixing and metallicity dependence of the stellar winds are accounted for (Meynet \& Maeder 2005). 
Models accounting for the effects of mass transfer in close binaries may also reproduce the observed trend (Eldridge et al. 2008). 
According to Foellmi et al. (2003ab) the proportion of detected binaries among WR stars in the
Magellanic Clouds is between 30-40\%. Taken at face, these numbers indicate that binarity might not be the dominant WR channel mechanism in these systems, especially when one considers the fact that
not all stars in binary systems undergo a Roche Lobe Overflow event (e.g. those binary systems in which the components are too far away from each other).
\item {\bf Populations of supergiants:} it is a well known fact that the number ratio of blue to red supergiant increases when the metallicity increases (Meylan \& Maeder 1982; Eggenberger et al. 2002). Standard models predict exactly the inverse behavior. As discussed in Langer \& Maeder (1995) this feature points probably toward
a missing mixing process in massive stars. It has been showed that in the SMC the observed blue to red supergiant ratio can
be reproduced by rotating models (Maeder \& Meynet 2001).
The blue to red supergiant ratio is also affected by mass loss. 
\item {\bf The frequency of different core-collapse supernovae:} the main types of core collapse supernovae are type II (H lines present in the spectrum), type Ib
(no H lines detected but He lines detected) and type Ic (no H, no He-lines detected). Many different subtypes are defined into these three categories (see e.g. Cappellaro et al. 2001). The frequency of the type Ibc supernovae normalized to the type II supernovae depends on the minimum initial mass required for stars to end their lifetime with no H-rich envelope. This is correct as long as the star formation rate can be considered constant in the last 50 million years and provided the evolution of single stars is mainly responsible for the value of this ratio.
Prantzos \& Boissier (2003) show that the above ratio increases with the metallicity. This has recently been confirmed by new data collected by Prieto et al. (2008). Models should explain such trend. We shall come back on that topic in Sect.~4 below.
\end{itemize}
Many other observational features might be added to the list above like the number ratios of
Be stars (fast rotating stars presenting an expanding equatorial disk) to B-type stars which increases
when the metallicity decreases (Maeder et al. 1999; Wisniewski \& Bjorkman 2006), the rotation rate of young pulsars (see the discussion in Heger et al. 2005), the shape of fast rotating stars as deduced from interferometry (see e.g. Carciofi et al. 2008), the variation with the latitude of the effective temperature at the surface of fast rotating stars (see e.g. Monnier et al. 2007), the wind anisotropies observed
for LBV stars and for Be stars (see Weigelt et al. 2007; Meilland et al. 2007ab), the ratio of the angular velocity of
the core to that of the surface and the size of the convective core obtained through
asteroseismology (see e.g. Aerts 2008)... Ideal stellar models should be able to reproduce satisfactorily all these important constraints. This is the price to pay to obtain reliable chemical yields, quantities of energy and momentum injected by massive stars. 
In the rest of this paper, we shall discuss two topics: the effects of rotation on the yields of CNO elements at low metallicity and the effect of rotation and mass loss on the properties of supernova progenitors.

\section{Early chemical evolution of the Milky Way}

Very metal poor halo stars
 have formed (at least in part) from matter  ejected  by very metal poor massive stars
 (see e.g. Chiappini et al 2005, 2008). Their surface
 compositions thus reflect
 the nucleosynthesis occurring in the first generations of massive stars
 (provided of course that no other processes as accretion or in-situ mixing
  mechanism has changed their surface composition).

\begin{figure}[t]
\begin{center}
 \includegraphics[width=3.5in,angle=0]{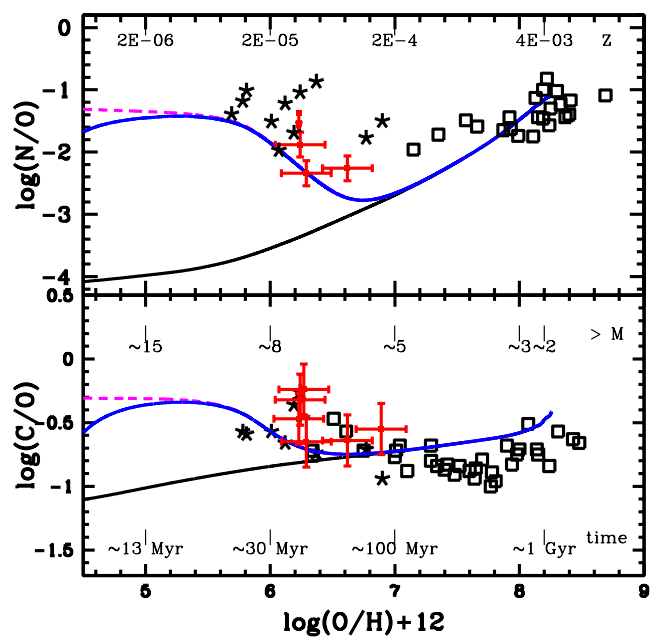} 
 \hfill
 \caption{Variation of the N/O and C/O ratios as a function the O/H ratios. Data points for halo stars are from Israelian et al.
 (2004, open squares ) and of Spite et al. (2005, stars). The points with error bars are for DLA systems from
 Pettini et al. (2008). The lower continuous curve is the chemical evolution model obtained with the stellar yields of slow rotating $Z=10^{-5}$ models from Meynet \& Maeder (2002) and Hirschi et al. (2004). The dashed line includes the yields of fast rotating $Z=10^{-8}$ models from Hirschi (2007) at very low metallicity. The intermediate curve is obtained using the yields of the $Z=0$ models presented in Ekstr\"om et al. (2008) up to $Z=10^{-10}$. The chemical evolution models are from
 Chiappini et al. (2006a), see Table~1 for the initial velocities of the stellar models.}
   \label{NOCO}
\end{center}
\end{figure}

 Interestingly, many observations of these stars show puzzling features.
 Among them let us cite the two following ones shown in Fig.~\ref{NOCO}: 1.-spectroscopic observations (e.g. Spite et al. 2005) indicate a primary production
 of nitrogen over a large metallicity range; 2.-
 halo stars with log(O/H)+12 inferior to about 6.5 present higher C/O ratios than
 halo stars with log(O/H)+12 between 6.5 and 8.2 (Akerman et al. 2004; Spite et al. 2005).

 Fast rotating
 massive stars are very interesting candidates
 for producing primary nitrogen at low metallicity. Their short lifetimes, together with their
 ability, when rotating sufficiently fast, to be important sources of primary nitrogen,
 allow them to account for the high observed N/O ratio at very low O/H
 values. Moreover these stars can also reproduce the observed 
 C/O upturn mentioned just above. This is illustrated in Fig.~\ref{NOCO}, where
 predictions for the evolution of N/O and C/O of chemical evolution models using different sets of yields are compared (Chiappini et al. 2006a\footnote{The details of the chemical evolution models can be found in Chiappini et al. (2006b), where they show that such a model reproduces nicely the metallicity distribution of the Galactic halo. This means that the timescale for the enrichment of the medium is well fitted.}). We see that the observed N/O ratio is much higher than what is predicted by a chemical evolution model using the yields of the slow-rotating $Z=10^{-5}$ models from Meynet \& Maeder (2002) down to $Z=0$. When adding the yields of the fast-rotating $Z=10^{-8}$ models from Hirschi (2007)
the fit is much improved. The same improvement is found for the C/O ratio, which presents an upturn at low metallicity. Thus these comparisons support fast rotating massive stars as the sources
of primary nitrogen in the galactic halo.
 
High N/O and the C/O upturn of the low-metallicity stars are also observed in low-metallicity DLAs  (Pettini et al. 2007, see the crosses in Fig.~\ref{NOCO}). We note that the observed points are
below the points for the halo stars in the N/O versus O/H plane. This may be attributed to two causes: either the observed N/O ratios observed in halo stars are somewhat overestimated or
the difference is real and might be due to different star formation histories in the halo and in DLAs. Let us just discuss these two possibilities. 

Measures of nitrogen abundances at the surface of very metal poor stars is quite challenging, much more than the measure of nitrogen in the interstellar medium as is done for the DLAs, therefore one expects that the data for DLAs suffer much smaller uncertainties than those for halo stars. 
In that respect 
the observed N/O ratios
in DLAs give more accurate abundances than halo stars. 
Most probably the star formation history in DLAs is not the same as in the halo. While, as recalled above, in the halo we see the result of a strong and rapid star formation episode, in DLAs one might see the result of  much slower and weaker star formation episodes. In that case, both massive stars and intermediate mass stars contributed to the build up of the chemical abundances and the chemical evolution models presented in Fig.~\ref{NOCO} no longer apply to these systems (see Chiappini et al. 2003; Dessauges-Zavadsky 2007 for chemical evolution models of DLAs). It will be very interesting to study the results of chemical evolution models adapted to this situation and accounting for stellar yields from both rotating massive and intermediate stars. Let us just mention at this stage that primary nitrogen production in metal poor intermediate mass stars is also strongly
favored when rotational mixing is accounted for (Meynet \& Maeder 2002). Thus also in that case, rotation may play a key role.

The primary nitrogen production is accompanied by other interesting features such as the production
of primary $^{13}$C (see Chiappini et al. 2008), and of primary $^{22}$Ne. Primary $^{22}$Ne is produced by diffusion of primary nitrogen from the H-burning shell to the core He-burning zone, or by the engulfment of part of the H-burning shell by the growing He-burning core. These processes
occur in rotating massive star models (Meynet \& Maeder 2002; Hirschi 2007). In the He-burning zone, $^{14}$N
is transformed into $^{22}$Ne through the classical reaction chain
$^{14}$N($\alpha$,$\gamma$)$^{18}$F($\beta^+$ $\nu$)$^{18}$O($\alpha$,$\gamma$)$^{22}$Ne.

\begin{figure}[t]
\begin{center}
 \includegraphics[width=2.5in,angle=-90]{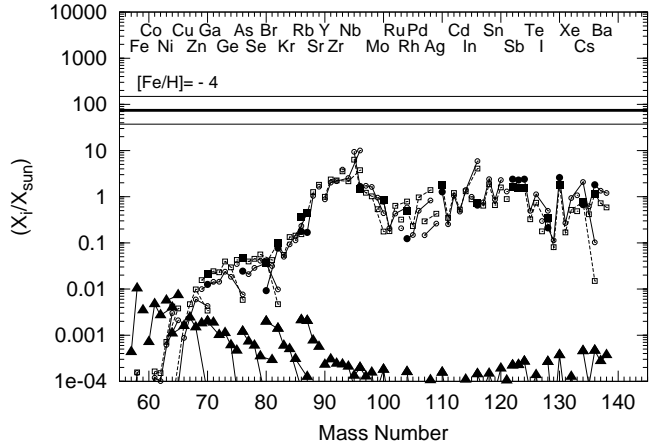} 
 \caption{s-process distributions between $^{57}$Fe and $^{138}$Ba normalized to solar for the 25 M$_\odot$ and [Fe/H]=-4 at the end of the convective C-burning shell. The horizontal line corresponds to the $^{16}$O overabundance in the C shell (thick line), multiplied and divided by two (thin lines). Isotopes of the same element are connected by a line. The cases presented are the following: {\it i)} non-rotating model (black triangles) and X($^{22}$Ne)$_{\rm ini}=5.21 \times 10^{-5}$; {\it ii)} rotating model (open squares, full squares for the s-only isotopes) and X($^{22}$Ne)$_{\rm ini}=5.0 \times 10^{-3}$; {\it iii)} rotating model (open circles, full circles for the s-only isotopes) and X($^{22}$Ne)$_{\rm ini}=1.0 \times 10^{-2}$. Figure from Pignatari et al. (2008).}
   \label{spro}
\end{center}
\end{figure}

In the He-burning zones (either in the core at the end of the core He-burning phase or in the
He-burning shell during the
core C-burning phase and in the following convective C-burning shell), neutrons are released through the reaction
$^{22}$Ne($\alpha$,n)$^{25}$Mg. 
These neutrons then can either be captured by iron seeds and produce s-process elements or be captured by light neutron poisons and
thus be removed from the flux of neutrons which is useful for s-process element nucleosynthesis.
The final outputs of s-process elements will
depend on at least three factors: the amounts of 1.- $^{22}$Ne, 2.- neutron poisons and 3.- iron seeds. In standard models (without rotation), when the metallicity decreases, the amount of $^{22}$Ne decreases
(less neutrons produced), the strength of primary neutron poisons becomes relatively more important in particular for [Fe/H]$\le$-2 with respect to solar, and the amount of iron seeds also decreases (e.g., Raiteri et al. 1992).
Thus very small quantities of s-process elements are expected (see the triangles in Fig.~\ref{spro}). 
When primary nitrogen and therefore primary $^{22}$Ne is present in quantities as given by rotating models
which can reproduce the observed trends for the N/O and C/O ratios in the
halo stars, then a very different output is obtained. 
The abundances of several s-process elements are increased by many orders of magnitudes. In particular, the elements are produced in the greatest quantities in the atomic mass region between strontium and barium,
and no long in the atomic mass region between iron and strontium as in the case of standard models.

These first results need to be extended for other masses, rotation and metallicities. However they
already show that some heavy s-process elements, not produced in standard models (without rotation), might be produced
in significant quantities in metal poor rotating stellar models. It will be very interesting in the future to find some non ambiguous signature of the occurrence of this process in the 
abundance pattern of very metal poor halo stars.

\section{Type Ib and Ic supernovae}

Core collapse supernovae of type Ib and Ic are very interesting events for many reasons. One of them is that in four cases, the typical spectrum of a type Ic supernova has been observed together with a long soft Gamma Ray Burst (GRB) event (Woosley \& Bloom 2006). Also, recent observations 
(Prieto et al. 2008) present new values for the variation with the metallicity of the number ratio 
(SN Ib+SN Ic)/SN II to which theoretical predictions can be compared. Finally, according at least to single star scenarios, these supernovae arise from the most massive stars. They offer thus a unique opportunity to study the final stages of these objects
which have a deep impact on the photometric and spectroscopic evolution of galaxies and also contribute to its chemical evolution .

We shall now discuss the predictions of single star models for the type Ib/Ic supernovae frequency.
Since these supernovae do not show any H-lines in their spectrum, they should have as progenitors stars having removed {\it at least} their H-rich envelope by stellar winds, {\it i.e.} their progenitors should be WR stars of the WNE type (stars with no H at their surface and presenting He and N lines) or of the WC/WO type.

\begin{figure}[t]
\begin{center}
 \includegraphics[width=2.8in,angle=-90]{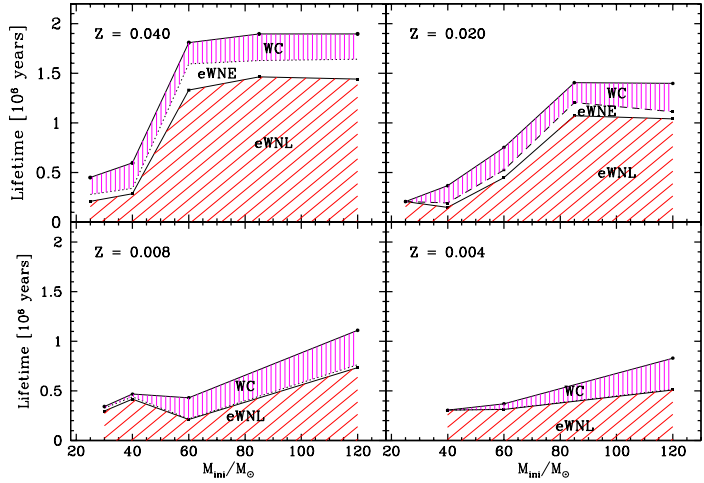} 
 \caption{Variation of the durations of the WR subphases as a function of the initial mass
at various metallicities. All the models begin their evolution with $\upsilon_{\rm ini}$ = 300 km s$^{-1}$. Figure taken from Meynet \& Maeder (2005). The small ``e'' in front of the different WR subtype indicates that criteria based on evolutionary computations have been used to classify the stars. In the text we have dropped the ``e'' and e.g. WNL=eWNL.}
   \label{tmatot}
\end{center}
\end{figure}

In Fig.~\ref{tmatot}, the duration of the different WR subphases is plotted as a function of the initial mass for
various 
metallicities. Only the results from models with rotation are plotted.
The greatest part of the WR lifetime is spent in the WNL phase (WR with still H in the envelope). Rotation increases the duration of this phase by allowing stars to enter this phase at an earlier stage of its evolution. 
For identical initial rotational velocities, the duration of the WNL phase is greater at higher metallicity. At higher metallicity, 
the higher mass loss rates by stellar winds enable the star to enter the WR phase at an earlier
stage. The WNE phase is also longer at higher metallicity (as is also the case for non--rotating models).
The WC phase keeps more or less
the same duration for all the metallicities in the higher mass star range. In the lower mass star
range the WC phase is longer at higher metallicity as a result of the shift toward a lower value of the minimum
initial mass of single stars needed to become a Wolf--Rayet star.

\begin{figure}[t]
\includegraphics[width=2.6in,height=2.6in,angle=0]{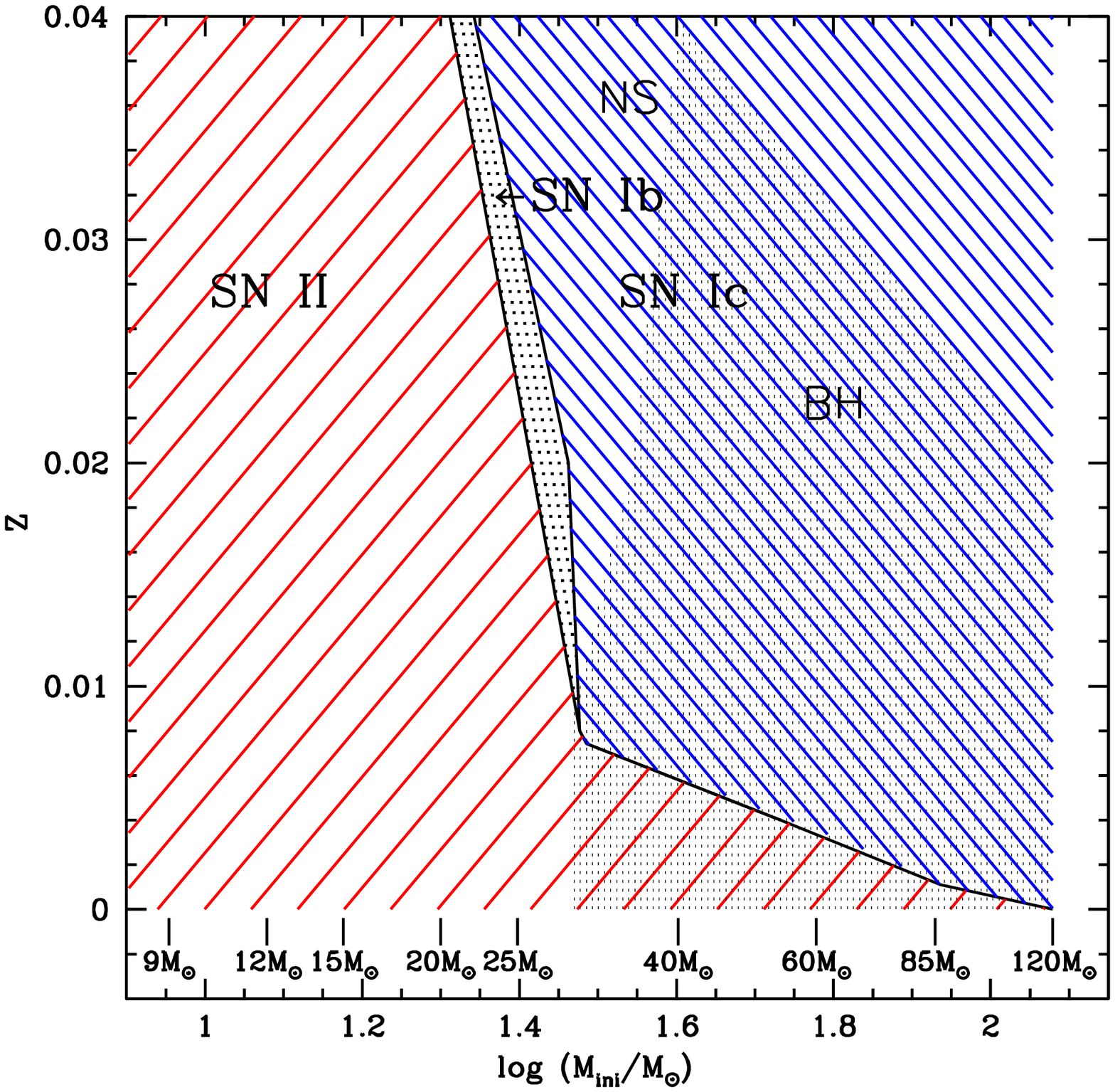}
\hfill
\includegraphics[width=2.6in,height=2.6in,angle=0]{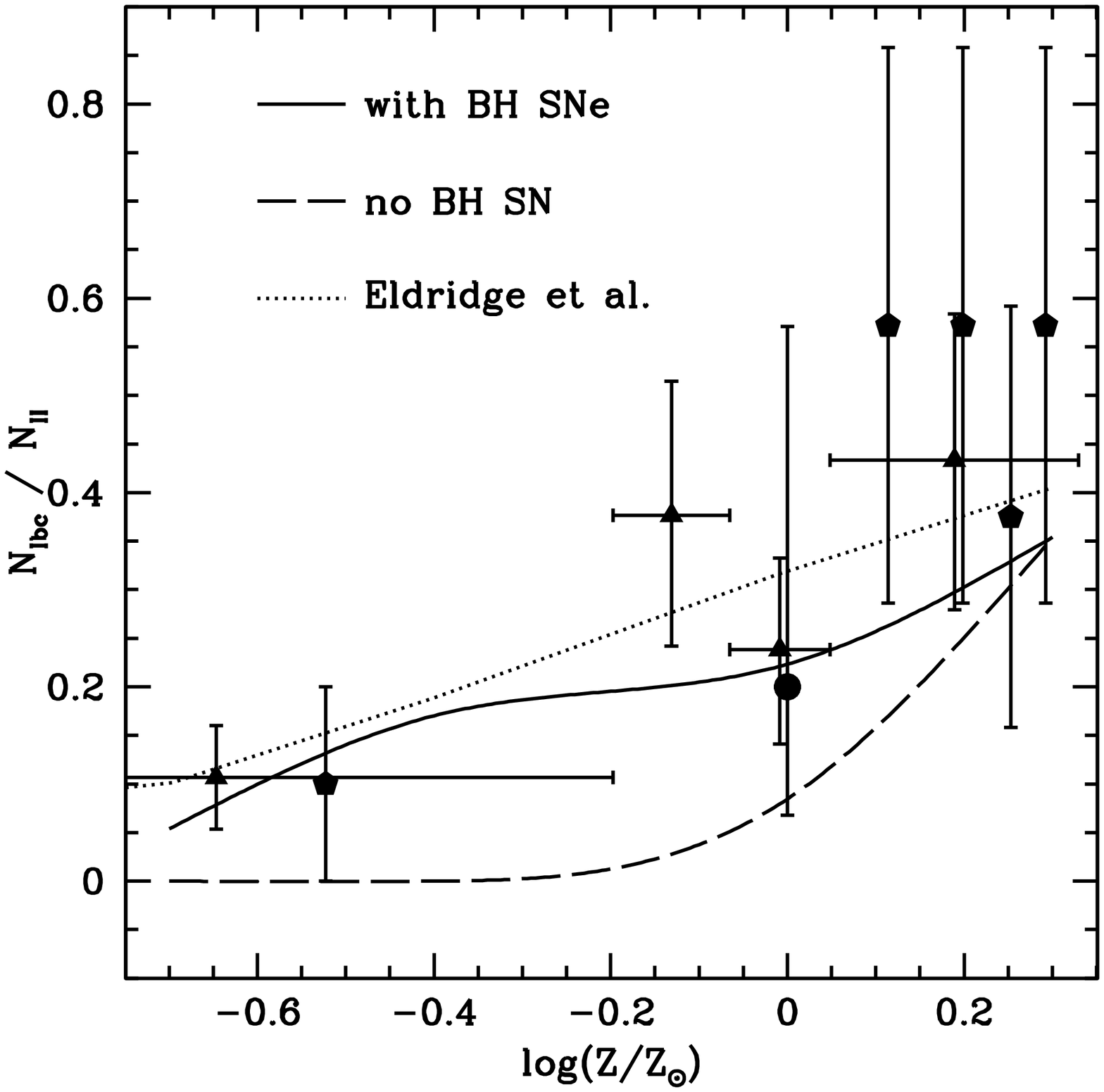}
\caption{{\it Left panel :}Ranges of masses of various types of SNe.
The area where formation of BH is expected is indicated as a superposed gray
region. {\it Right panel :} Rate of SN Ibc / SN II if all models produce a SN (solid line) or if models producing a black hole do not explode in a SN (dashed line). Pentagons are observational data from Prieto et al. (2008), and triangles are data from Prantzos \& Boissier (2003). The dotted line represents the binary models of Eldridge et al. (2008).}
\label{WR}
\end{figure}

To link these results with the type of the supernova event we adopted the following rules:
as long as some hydrogen is present in the ejecta, a type II supernova will occur. 
We considered that all supernovae ejecting less than about $0.6 \,\mathrm{M}_\odot$ of helium are of type Ic. All progenitors satisfying neither the criterion for becoming a type II (no H present
in the ejecta), neither the one for becoming a type Ib (more than $\sim$0.6 M$_\odot$ of He) are considered to give rise to type Ib SN events. 
Fig.~\ref{WR} shows
the different supernova types expected for various initial masses and metallicities.
Looking at that figure, we can make the following remarks:
\begin{itemize}
\item As expected the lower mass limit for having a type Ibc SN decreases when the metallicity increases. We see that the dependence on the metallicity of this limit is much stronger at low Z
than at high Z;
\item The mass range of stars ending their life in a type Ib supernova event is relatively narrow. Below this mass range, the stars do not succeed in entering the WR phase, above, mass loss rates are efficient enough for allowing the star to further evolve into the WC or WO phase.
\item Above 30 M$_\odot$ and for metallicities higher than 0.008, all stars end their life
as type Ic SNe. For metallicities below 0.008, the minimum initial mass for stars ending their
lifetime as type Ic rapidly increases. 
\end{itemize}

The mass range for type Ic supernovae is thus much larger than the mass range for type Ib SNe.
Even when weighted with a Salpeter IMF, one expects thus that the frequency of type Ic SNe is higher than that of type Ib.
Interestingly also, one can note from the mass of helium ejected at the time of the SN event, that
the minimum amount of He ejected by type Ic event is 0.3 M$_\odot$.

Considering that all models ending their lifetime as a WNE or WC/WO phase will explode as a type Ibc supernova, it is possible to compute the variation with the metallicity of the number ratio of type Ibc to type II supernovae. The result is shown in Fig.~\ref{WR} (right panel). One sees that
this ratio increases with the metallicity. This is due to the fact that at higher metallicity, the minimum initial mass of stars ending their life as WNE, or WC/WO stars is lower than at lower
metallicities. Single star models can reasonably well reproduce the observed trend with the metallicity. They however give slightly too small values with respect to the observations, which may indicate that a portion of the type Ibc supernovae may originate from close binary evolutions. 
Models accounting
for single and binary channel (but without rotation) are shown as a dotted line (Eldridge et al. 2008). They provide a good
fit to the observations. But in that case most of the supernovae originate from the binary channel,
leaving little place for the single star scenario. These models would also predict that most of the
WR stars are the outcome of close binary evolution. This does not appear to be confirmed by the
observations of Foellmi et al (2003ab, see above).
Most likely, both the single and binary channel contribute.

The results of single star models may change if, when a Black Hole (BH) is formed, no SN event occurs. 
Superposed to Figs.~\ref{WR} (left panel), we have indicated, as a light gray dotted zone, the
regions where a black hole might be formed instead of a neutron star. For drawing this zone, we
assumed that $2.7\,\mathrm{M}_\odot$ as the maximum mass of a neutron star. This value is compatible with the one given by Shapiro \& Teukolsky (1983) and also with the recent discovery of a massive neutron star of $2.1\,\mathrm{M}_\odot$ (Freire et al. 2008). Adopting this mass limit and the relation $M_{\rm NS}$ versus $M_{\rm CO}$ deduced from the models of Hirschi et al. (2005), we can estimate for each metallicity the mass ranges of stars producing neutron stars, respectively black holes. 
At low metallicity (Z $\le$ 0.01), all stars with an initial mass larger than $30\,\mathrm{M}_\odot$ finish their life as a BH, that is, all WR stars and the most massive red-- or blue--supergiants. 
The mass range of BH progenitors is decreasing when more and more metal rich environments are considered. For $Z \ge \sim 0.040$, no BH is expected from single star scenario.
(see Fig.~\ref{WR}). This is of course an effect due to increased mass loss at higher metallicities. The stars lose too much mass for forming BHs.

We note that in the frame of our hypothesis, all the type Ib SNe give birth to a neutron star.
For $Z \ge \sim 0.03$, most of the type Ic supernovae also produce a neutron star. Below a metallicity of 0.008, all WR progenitors produce a black hole.
If, when a BH is formed no supernova event is obtained, then one expects no type Ibc supernovae (from single stars) below a metallicity of about 0.008.

We computed new (SN Ib + SN Ic)/ SN II ratios 
with the assumption that all models massive enough to form a black hole do not produce a SN. 
Comparing with the observed rates in Fig.~\ref{WR} (see dashed line in right panel)
we see that in the case no supernova event occurs when a BH is formed, 
single star models might still account for a significant fraction of the type Ibc supernovae
for $Z > 0.02$.
At $Z=0.004$ all type Ibc should arise from other evolutionary scenarios, probably involving close binary evolution with mass transfer. 
If we take a smaller maximal mass for neutron stars, this situation is still more extreme: almost all the WR stars produce a black hole, thus the rate of SN Ibc / SN II is null or very small at each metallicity. 

Probably, the hypothesis according to which no supernova event is associated when a BH is formed, is too restrictive. For instance, 
the collapsar scenario for Gamma Ray Bursts (Woosley 1993)  needs the formation of a black holes (Dessart, private communication) and this formation is at least accompanied in some cases by a type Ic supernova event. 
Also the observation of 
the binary system GRO J1655-40 containing a black hole (Israelian et al. 1999) suggest that a few stellar masses have been ejected  and therefore a SN event occurred when the BH formed. This is deduced from the important chemical anomalies observed at the surface of the visible companion, chemical anomalies whose origin is attributed to the fact that the (now visible) companion accreted part of the SN ejecta. This gives some
support to the view that, at least in some cases, the collapse to a BH does not prevent a supernova event to occur.

To conclude, we see that still major improvements of massive star stellar models are still needed in order to improve our understanding of their impacts in galaxies. Probably a key step will be made forward when  we shall also have a better understanding on how  massive stars form. This is crucial to link stellar physics to the physics of galaxies.


\end{document}